\documentclass[reprint,
superscriptaddress,
bibnotes,
 amsmath,amssymb,
 aps,
pra,
showkeys,
floatfix,
]{revtex4-1}

\usepackage{graphicx}
\usepackage{dcolumn}
\usepackage[T1]{fontenc}
\usepackage[utf8]{inputenc}
\usepackage{graphicx}
\usepackage{float}
\usepackage{natbib}
\usepackage{hyphenat}
\usepackage{tabularx}
\usepackage{amsmath}
\usepackage{adjustbox}
\usepackage{longtable}
\usepackage{lipsum}
\usepackage{epstopdf}
\DeclareUnicodeCharacter{0301}{\'{e}}
\usepackage{bm}

\begin{document}

\preprint{APS/123-QED}

\title{Simultaneous observation of anti-damping and inverse spin Hall effect in La$_{0.67}$Sr$_{0.33}$MnO$_{3}$/Pt bilayer system}

\author{Pushpendra Gupta}
\affiliation{Laboratory for Nanomagnetism and Magnetic Materials (LNMM), School of Physical Sciences, National Institute of Science Education and Research (NISER), HBNI, P.O.- Bhimpur Padanpur, Via –Jatni, 752050, India}

\author{Braj Bhusan Singh}
\affiliation{Laboratory for Nanomagnetism and Magnetic Materials (LNMM), School of Physical Sciences, National Institute of Science Education and Research (NISER), HBNI, P.O.- Bhimpur Padanpur, Via –Jatni, 752050, India}

\author{Koustuv Roy}
\affiliation{Laboratory for Nanomagnetism and Magnetic Materials (LNMM), School of Physical Sciences, National Institute of Science Education and Research (NISER), HBNI, P.O.- Bhimpur Padanpur, Via –Jatni, 752050, India}

\author{Anirban Sarkar}%
\affiliation{Forschungszentrum Jülich GmbH, Jülich Centre for Neutron Science (JCNS-2) and Peter Grünberg Institut (PGI-4), JARA-FIT, 52425 Jülich, Germany}

\author{Markus Waschk}
\affiliation{Forschungszentrum Jülich GmbH, Jülich Centre for Neutron Science (JCNS-2) and Peter Grünberg Institut (PGI-4), JARA-FIT, 52425 Jülich, Germany}

\author{Thomas Brueckel}
\affiliation{Forschungszentrum Jülich GmbH, Jülich Centre for Neutron Science (JCNS-2) and Peter Grünberg Institut (PGI-4), JARA-FIT, 52425 Jülich, Germany}

\author{Subhankar Bedanta}
\email{sbedanta@niser.ac.in}
\affiliation{Laboratory for Nanomagnetism and Magnetic Materials (LNMM), School of Physical Sciences, National Institute of Science Education and Research (NISER), HBNI, P.O.- Bhimpur Padanpur, Via –Jatni, 752050, India}
\date{\today}

\begin{abstract}
  Manganites have shown potential in spintronics because they exhibit high spin polarization. Here, by ferromagnetic resonance  we have studied the damping properties of La$_{0.67}$Sr$_{0.33}$MnO$_{3}$/Pt bilayers which are prepared by oxide molecular beam epitaxy. The damping coefficient ($\alpha$) of La$_{0.67}$Sr$_{0.33}$MnO$_{3}$ (LSMO) single layer is found to be 0.0104. However the LSMO/Pt bilayers exhibit decrease in $\alpha$ with increase in Pt thickness. This decrease in the value of $\alpha$ is probably due to high anti-damping like torque. Further, we have investigated the angle dependent inverse spin Hall effect (ISHE) to quantify the spin pumping voltage from other spin rectification effects such as anomalous Hall effect and anisotropic magnetoresistance. We have observed high spin pumping voltage ($\sim$~20 $ \mu  V$). The results indicate that both anti-damping and spin pumping phenomena are occuring simultaneously.
\end{abstract}

\maketitle

\section{\label{sec:level1}Introduction:\protect}
Spintronics devices have demonstrated high data storage capacity and miniaturization of computer logics. For the development of next generation devices low power and high speed are the key requisite. Pure spin current ($J_{s}$) based devices have shown potentials for fulfilling these requirements due to minimal involvement of charge current ($J_{c}$). In this context, ferromagnetic (FM)/heavy metal (HM) hetrostructures are model systems to investigate various spin dependent phenomenon \cite{wolf2001spintronics,chappert2010emergence,vzutic2004spintronics,pham2016ferromagnetic}.  

Generation of pure spin current has been demonstrated by ferromagnetic resonance (FMR) through spin pumping mechanism~\cite{mosendz2010quantifying,cerqueira2018evidence,tserkovnyak2002spin,jamali2015giant}. This pure spin current can lose their spin angular momentum in the presence of high spin orbit coupling (SOC) HM materials e.g Pt, W, Ta etc. The loss of spin angular momentum can develop a voltage by asymmetric scattering of spin, which is known as inverse spin Hall effect (ISHE)~\cite{hirsch1999spin,valenzuela2006direct}. SOC is an important factor for observation of large ISHE. Because of the spin-orbit interaction, different spins (up and down) move in different direction and hence an electric field is developed transverse to the direction of spins \cite{kimura2007room,saitoh2006conversion,takeuchi2008charge}.
In ISHE process  $J_{s}$ is converted to  $J_{c}$ and these two physical parameters are related by the below equation:
\begin{equation}
J_{c}=\theta_{SH}\times J_{s}\times \sigma
\label{eq1}
\end{equation} 
where $\theta_{SH}$ is the spin Hall angle (SHA) and $\sigma$ is the polarization vector transverse to the direction of $J_{s}$. The value of SHA, therefore, defines the charge to spin current conversion efficiency. The absorption of spin current ($J_{s}^{abs}$) generated by HM into FM create a spin transfer torque, which can be quantified by spin torque efficiency $\xi_{SH}$ = ($2e/\hbar$)$J_{s}^{abs}$/$J_{c}$ \cite{nguyen2015enhancement}. It is noted here that in such FM/HM heterostructures, spin pumping increases the value of $\alpha$ due to absorption of spin angular momentum in HM layer\cite{ando2008electric,mizukami2001ferromagnetic}. Further in such FM/HM bilayers another type of torque may occur which is called as anti-damping torque\cite{emori2015quantification,behera2017two}.This later torque will lead to a decrease in damping value of the bilayer as compared to the reference single FM layer. It is known that a large value of $\xi_{SH}$ and lower $\alpha$ are the important parameters for the development of power efficient devices. Therefore the anti-damping torque may help in achieving magnetization switching at lower current density which is proportional to $\alpha$/$\xi_{SH}$, where $\alpha$ is the damping constant of FM/HM bilayer\cite{nguyen2015enhancement}. We note that keeping low $\alpha$ value with spin pumping is  a challenge. However anti-damping like torque may help to balance the damping like torque which is opposite of that and hence reduce the value of $\alpha$ in FM/HM heterostructures.
Pt has been used widely due to its high conductivity and SHA values. Studies so far are concentrated mostly on Pt and ferromagnetic metals \cite{kimura2007room,rojas2014spin,huo2017spin,zheng2017large}. In this context insulating ferromagnetic oxides in particular manganites are worth to be investigated for spin to charge conversion based applications.   La$_{0.67}$Sr$_{0.33}$MnO$_{3}$ (LSMO) is one such ferromagnetic oxide well known for exhibiting high Curie temperature ($T_C$ $\sim$ 350\,K) and nearly 100\% spin polarization (in bulk)~\cite{park1998direct}. There are a few reports where spin pumping has been investigated in LSMO/Pt bilayers for which the LSMO is primarily prepared by pulsed laser deposition technique. \cite{luo2017spin,lee2016magnetic,luo2014influence,atsarkin2016resonance,luo2012spin,luo2015enhanced}   In this work, we aim to study LSMO/Pt bilayers where the samples have been fabricated by oxide molecular beam epitaxy (OMBE) technique. In recent years OMBE has been proven to be an excellent technique to grow high qualilty complex oxide thin films. Here we show that our LSMO/Pt films are highly resistive. Further we have observed high spin pumping voltage. Both these factors have led to a $\theta_{SHA}$ of 0.033.  We have also observed decrease in the value of $\alpha$ with increase in spin pumping voltage, which make them very useful for spintronics devices.

\section{\label{sec:level2}Experimental METHODS:\protect}
LSMO($t_{LSMO}$ = 20\,nm)/Pt($t_{Pt}$) bilayer samples have been prepared on single crystalline SrTiO$_{3}$(001) substrate using an oxygen plasma assisted molecular beam epitaxy (MBE) system. Samples are named as S1, S2 and S3 for the thickness of Pt ($t_{Pt}$) = 0, 3 and 10 nm, respectively. Surface and crystalline quality of LSMO films were characterized by $in$ $situ$ low energy electron diffraction (LEED) and reflective high energy electron diffraction (RHEED). X-ray diffraction was performed to determine the crystal phases. Film thicknesses were obtained using X-ray reflectivity. 
Magnetization dynamics was studied using co-planer wave-guide (CPW) based ferromagnetic resonance (FMR) spectroscopy. Sample was kept on top of CPW in a flip-chip manner~\cite{singh2017study} as shown in Fig.\ref{fig:figure-1}(a). To avoid shunting a 25 $\mu$m polymer tape was used between sample and CPW. A DC magnetic field $H$, perpendicular to radio frequency field ($h_{rf}$), was applied using an electromagnet. Gilbert damping constant ($\alpha$) was extracted by measuring FMR spectra in a frequency ($f$) range of 3-16 GHz with intervals of 0.5 GHz. The values of resonance field ($H_{res}$) and linewidth ($\Delta H$) have been obtained from the Lorentzian fit of the FMR spectra, while the  $\alpha$ has been evaluated by fitting the $\Delta H$  vs $f$ data.
ISHE voltage was measured by a nanovoltmeter. Detailed description of the instrument is described in our previous work ~\cite{singh2017study,singh2019inverse,PhysRevApplied.13.044020,nayak2018effect}. The measurements were performed on samples of dimension $\sim$ 3$\times$2 mm$^{2}$. Copper wires were used to make contacts using silver paste at the edges of the samples. Angle dependent ISHE has been performed at $f$ = 7 GHz, to disentangle spin rectification effects. Microwave power dependent ISHE measurement has been performed using $rf$ signal generator(SMB--100 model from ROHDE \& SCHWARZ).

\section{\label{sec:level3}Results and discussion:\protect}
\begin{figure}[ht]
	\centering
	\includegraphics[width=0.48\textwidth]{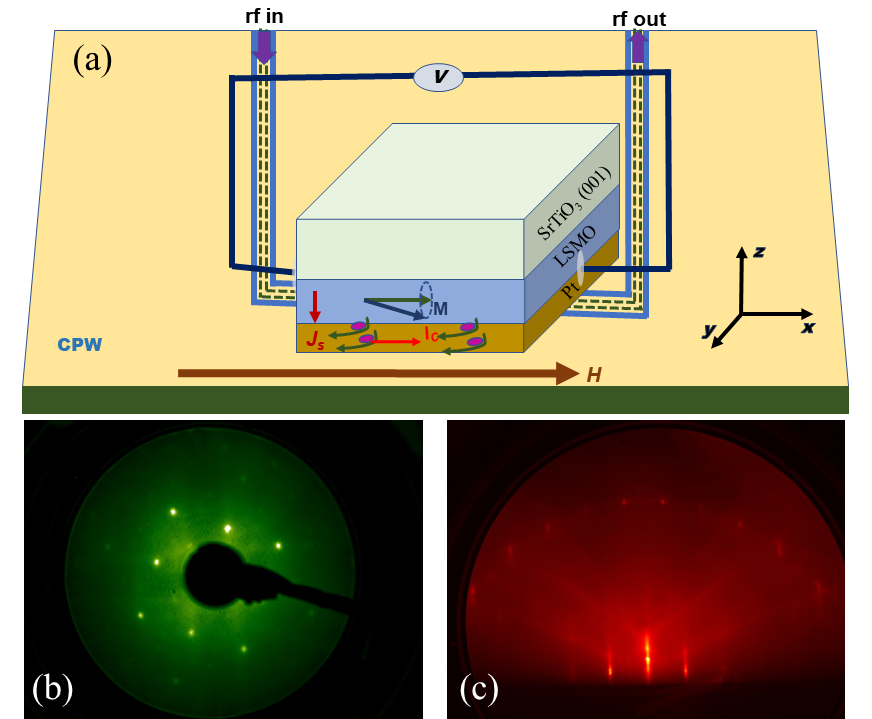}
	\caption{(a) Experimental setup for FMR and ISHE measurements. (b) LEED (c) RHEED images for sample S1.}
	\label{fig:figure-1}
\end{figure} 
Figure~\ref{fig:figure-1}(b) and (c) show the LEED and RHEED images for the 20 nm thick LSMO film (sample S1), respectively. The presence of sharp spots (Fig. 1(b)) and streaks (Fig. 1(c)) confirms  epitaxial growth of LSMO films on the SrTiO$_{3}$(001) substrate. The RHEED image also indicate a smooth surface of the LSMO film. These were also confirmed by the x-ray diffraction (data shown in supplementary information).

Figure~\ref{fig:figure-2}(a) show the $f$ vs $H_{res}$ plot for all the samples obtained from the frequency dependent FMR spectra. The data have been fitted by Kittle equation~\cite{kittel1948theory},
\begin{equation}
f=\frac{\gamma}{2\pi}\sqrt{(H_{res}+H_{K})(H_{res}+4\pi M_{eff}+H_{K})}
\label{kittle equation}
\end{equation}
where $\gamma$(=$\frac{g\mu_{B}}{\hbar}$), $g$, $\mu_{B}$ and $M_{eff}$ are gyromagnetic ratio, Lande g--factor, Bohr magnetron and effective magnetization, respectively. $H_{K}$, $K_{S}$ and $t_{FM}$ are anisotropic field, perpendicular surface anisotropy constant, and thickness of the LSMO layer, respectively. Further $\alpha$ was evaluated by fitting data of Fig.~\ref{fig:figure-2}(b) using the relation 
\begin{equation}
\Delta H=\Delta H_{0}+\dfrac{4\pi\alpha f}{\gamma} .
\label{non liner fit}
\end{equation} 

\begin{figure}[b]
	\centering
	\includegraphics[width=0.48\textwidth]{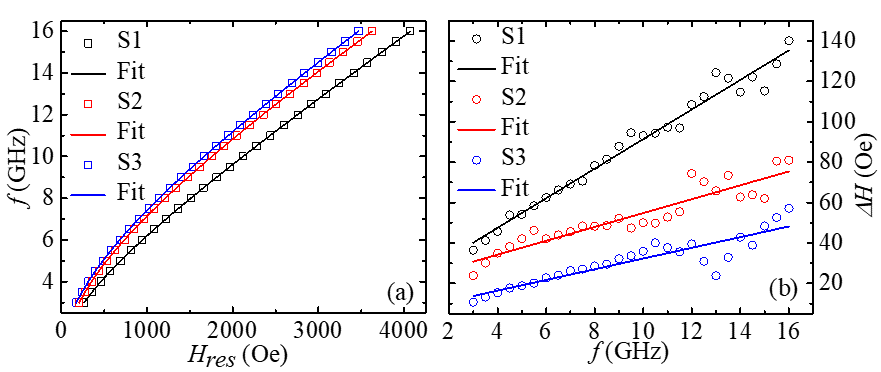}
	\caption{(a) $f$ vs $H_{res}$ and (b) $\Delta H$ vs $f$ for samples S1, S2 and S3.}
	\label{fig:figure-2}
\end{figure}

The values of $\alpha$ for samples S1, S2 and S3 are extracted to be 0.0104$\pm$0.0003, 0.0046$\pm$0.0004 and 0.0037$\pm$0.0004, respectively. It should be noted that Pt is a well known metal for exhibiting high SOC and when coupled to a FM layer it may lead to an increase in $\alpha$. However in our case it is observed that there is a decrease in $\alpha$ with increase in $t_{Pt}$ in comparison to the single LSMO layer (S1). The reason for this lowering of $\alpha$ could be an anti-damping like torque. 
Similar anti-damping behavior has been observed in $\beta$-Ta and Py bilayer system studied by Behera \textit{et. al.}~\cite{behera2016anomalous}.
The anti-damping in a FM/HM heterostructure can be explained in the following manner. In case of spin flip parameter ($\varepsilon$)$<$0.1, spin angular momentum at the FM/HM interface creates a non-equilibrium spin density in the Pt layer~\cite{tserkovnyak2005nonlocal}. This results a back flow of spin current ($J_{s}^{0}$) into the LSMO layer which has two components, (i) parallel, and (ii) perpendicular to instantaneous magnetization $m(t)$ of LSMO layer. The parallel component to $m(t)$ counteracts the spin pumping from LSMO layer and suppresses the spin pumping in Pt layer. Component which is transverse to $J_{s}^{0}$ generates an additional spin orbit torque (SOT) on this in-plane $m(t)$ of LSMO layer. This SOT can be effective up to a distance twice of the spin diffusion length ($\lambda_{SD}$)~\cite{jiao2013spin}. 
Spin accumulation at the interface is very sensitive to $\lambda_{SD}$ of Pt layer. For $t_{Pt}$ $<$ $\lambda_{SD}$, spin accumulation dominates over the bulk SOC of Pt. This leads to an increase in $J_{s}^{0}$, and a decrease of $\alpha$. For $t_{Pt}$ $>$ $2\lambda_{SD}$, $J_{s}^{0}$  decreases which results in a decrease of SOT. This decrease in SOT may lead to an increase in $\alpha$. 
We have considered $\lambda_{SD}$ $\sim$ 5.9 nm from literature, where the samples studied had similar type of structure~\cite{luo2017spin}. Therefore, it can be concluded that anti-damping like torque is very high and opposite in our samples to overcome damping like torque, which leads to the reduction of the value of $\alpha$.
\begin{figure}[ht]
	\centering
	\includegraphics[width=0.48\textwidth]{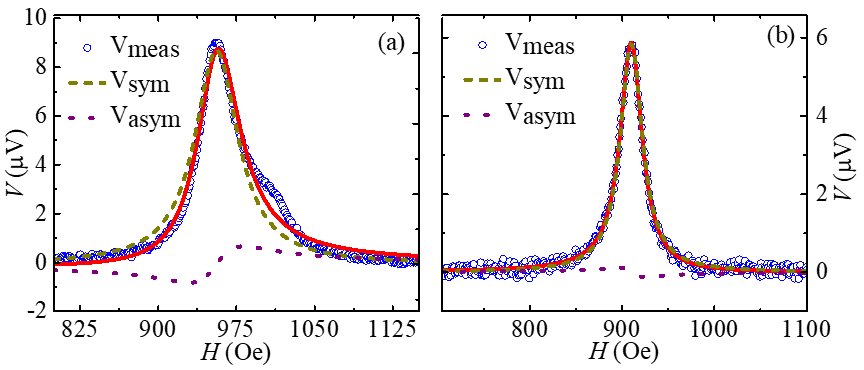}
	\caption{ISHE voltage for samples S2 and S3 are shown in (a) and (b), respectively. Open circles (in blue) is the measured ISHE voltage and solid line (in red) represents the best fit of the data fitted by equation (4). Dash (in green)  and dot (purple) lines represent the $V_{sym}$ and $V_{asym}$ components, resepctively, evaluated by fitting to equation (4).}
	\label{fig:figure-3}
\end{figure}
In order to quantify the spin pumping we have performed ISHE measurements. Figure~\ref{fig:figure-3}(a) and (b) represent the ISHE voltage ($V_{meas}$) vs. applied field ($H$) for samples S2 and S3, respectively. It is noted that no ISHE signal has been observed for the reference sample S1. We have separated symmetric ($V_{sym}$) and anti-symmetric ($V_{asym}$) voltage signal by using the following equation~\cite{iguchi2016measurement},
\begin{multline}
V_{total}=V_{sym}\frac{(\Delta H)^{2}}{(H-H_{res})^{2}+(\Delta H)^{2}}+\\
V_{asym}\frac{(\Delta H)(H-H_{res})}{(H-H_{res})^{2}+(\Delta H)^{2}} .
\end{multline}
 Fig.~\ref{fig:figure-3}(a) and 3(b) show that  $V_{sym}$ component is large in comparison to $V_{asym}$. It is well known that symmetric signal comes predominantly from the spin pumping while asymmetric signal is due to other rectification effects~\cite{iguchi2016measurement}. 
\begin{figure}[ht]
	\centering
	\includegraphics[width=0.48\textwidth]{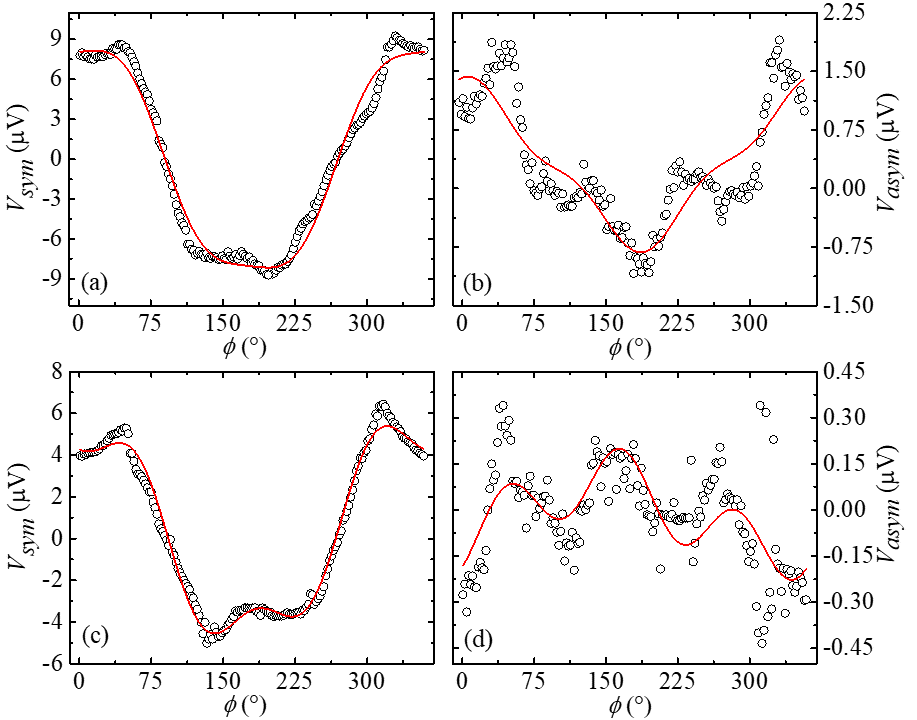}
	\caption{(a) and (c) Angle dependent $V_{sym}$ for samples S2 and S3, (b) and (d)angle dependent $V_{asym}$ for samples S2 and S3, Figure (a) and (c) were fitted by using equation (5) while figure (b) and (d) were fitted by using equation (6)}
	\label{fig:figure-4}
\end{figure}
To separate the effect of anomalous Hall effect (AHE) and anisotropic magneto resistance (AMR) from spin pumping we have performed angle $(\phi)$ dependent ISHE at a step of $2^{0}$ in the range of 0 to $360^{0}$. Here $(\phi)$ is defined as the  angle between direction of applied magnetic field ($H$) and the contacts for voltage  measurement. Figure~\ref{fig:figure-4}(a) and (b) show the angle dependent $V_{sym}$ and $V_{asym}$ for sample S2, respectively. Similarly, Fig.~\ref{fig:figure-4}(c) and (d) show the evaluated $V_{sym}$ and $V_{asym}$ for the sample S3. These plots were fitted using the following relations~\cite{conca2017lack},
\begin{multline}
V_{sym}=V_{sp}\ cos^3\phi+V_{AHE}\ cos\phi\ cos\theta+ \\ V_{sym}^{AMR\perp}\ cos(2\phi)  cos\phi+ V_{sym}^{AMR\parallel} \sin(2\phi)\ cos\phi
\end{multline}
\begin{multline}
   V_{asym}=V_{AHE}\ cos\phi\ sin\theta +\ V_{asym}^{AMR\perp}\ cos(2\phi)\ cos\phi +\\ V_{asym}^{AMR\parallel}\ sin(2\phi)\ cos\phi 
\end{multline}
where $V_{sp}$, $V_{AHE}$ are voltages due to spin pumping and anomalous Hall effect. Further $V_{asym,sym}^{AMR\parallel}$ and $V_{asym,sym}^{AMR\perp}$ are the parallel and perpendicular components of the AMR voltage, respectively. $\theta$ is the angle between $h_{rf}$ and $H$ which is 90$^\circ$ in our case as shown in figure 1(a)\ref{fig:figure-1}. So the equations (5) and (6) can be written as
\begin{multline}
V_{sym}=V_{sp}\ cos^3\phi+ V_{sym}^{AMR\perp}\ cos(2\phi) \ cos\phi+ \\V_{sym}^{AMR\parallel} sin(2\phi)\  cos\phi
\end{multline}
\begin{multline}
V_{asym}=V_{AHE}\  cos\phi+V_{asym}^{AMR\perp} cos(2\phi)\ cos\phi +\\ V_{asym}^{AMR\parallel} sin(2\phi) \ cos\phi 
\end{multline}
$V_{AMR} ^{\perp,||}$ can be evaluated by the following equation ~\cite{conca2017lack}
\begin{equation}\label{q11}
V_{AMR} ^{\perp,||}=\sqrt{(V_{asym}^{AMR \perp,||})^{2}+(V_{sym}^{AMR \perp,||})^{2}}  
\end{equation}
The values $V_{sp}$, $V_{AHE}$, $V_{AMR}^{\perp}$ and $V_{AMR}^{\parallel}$ for samples S2 and S3 were obtained from the best fits and listed in table I. 
\begin{table}[H]
\caption{Fitted parameter for samples S2 and S3} \label{tab:title} 
\begin{center}
	\begin{tabular}{c c c c c} 
		\hline
		Sample & $V_{sp}(\mu V)$ & $V_{AHE}(\mu V)$ & $V_{AMR}^{\perp}(\mu V)$ & $V_{AMR}^{\parallel}(\mu V)$ \\ [1ex] 
		\hline
		S2 & 20.05$\pm$0.28 & 0.77$\pm$0.05 & 11.98$\pm$0.36 & 0.34$\pm$0.22 \\ 
		\hline
		S3 & 12.79$\pm$0.11 & -0.01$\pm$0.01 & 8.49$\pm$0.37 & 0.55$\pm$0.07 \\
		\hline		
	\end{tabular}
\end{center}
\end{table}

From Table I it is observed that $V_{sp}$ decreases for higher Pt thickness. 

 We have calculated SHA by using below equation (10) ~\cite{rogdakis2019spin,jeon2018spin}.
\begin{multline}
 V_{ISHE}=(\frac{w}{t_{LSMO}/\rho_{LSMO} + t_{Pt}/\rho_{Pt}})\times\\ \theta_{SHA} \lambda_{SD}tanh[\frac{t_{Pt}}{2 \lambda_{SD}}] J_s
\end{multline}
where $J_{s}$ is given by,
\begin{multline}
J_s \approx (\frac{g_{r}^{\uparrow \downarrow}\hbar}{8\pi})(\frac{\mu_0 h_{rf}\gamma}{\alpha})^2\times\\
[\frac{\mu_0 M_s\gamma+\sqrt{(\mu_0 M_s\gamma)^2+16(\pi f)^2}}{(\mu_0 M_s\gamma)^2+16(\pi f)^2}](\frac{2e}{\hbar})
\end{multline}
and
\begin{flalign}
g_{r}^{\uparrow\downarrow}=g_{eff}^{\uparrow\downarrow}[1+\dfrac{g_{eff}^{\uparrow\downarrow}\rho_{Pt}\lambda_{SD}e^2}{2\pi\hbar\tanh[\dfrac{t_{Pt}}{\lambda_{SD}}]}]^{-1} 
\end{flalign}
where $w$, $M_{s}$, and $g_{eff}^{\uparrow\downarrow}$ are the width of CPW, saturation magnetization and spin mixing conductance of the bilayers, respectively. For the evaluation of $g_{eff}^{\uparrow\downarrow}$ the resistivity ($\rho$) of the samples were calculated by four-probe method. The ($\rho$) values are $4.79\times 10^{-5}$, $7.33\times 10^{-7}$ and $5.25\times 10^{-7}$ $\Omega$-m for the samples S1, S2 and S3, respectively.
The value of $g_{\it{eff}}^{\uparrow \downarrow}$ can be calculated by the following expression using damping constant \cite{saitoh2006conversion}:
\begin{equation}\label{q12}
g_{eff}^{\uparrow \downarrow}=\frac{\Delta\alpha 4\pi M_{s}t_{LSMO}}{g\mu_{B}} 
\end{equation}
where $\Delta\alpha$ is the change in the $\alpha$ due to spin pumping. The values of $\theta_{SHA}$ are evaluated to be 0.033 and 0.014 for samples S2 and S3, respectively. These $\theta_{SHA}$ values well matched to the previously reported values for the Pt~\cite{wang2014determination,tao2018self,wang2016giant} in a similar type of system. For comparison to LSMO/Pt system we have also prepared one sample S4 with structure  Si/Co$_{20}$Fe$_{60}$B$_{20}$ (5 nm)/Pt (3 nm) sample by DC sputtering system. Detailed analysis has been shown in supplementary information for this sample. The calculated value of $\theta_{SHA}$ for sample S4 is 0.022, which is in range of $\theta_{SHA}$ for Pt in LSMO/Pt system as mentioned earlier. It is to be noted that $V_{sp}$ for sample S2 is nearly 15 times higher than sample S4, however the enhancement in $\theta_{SHA}$ for S2 in comparison to S4 is only 1.1\%.

\begin{figure}[t]
	\centering
	\includegraphics[width=0.48\textwidth]{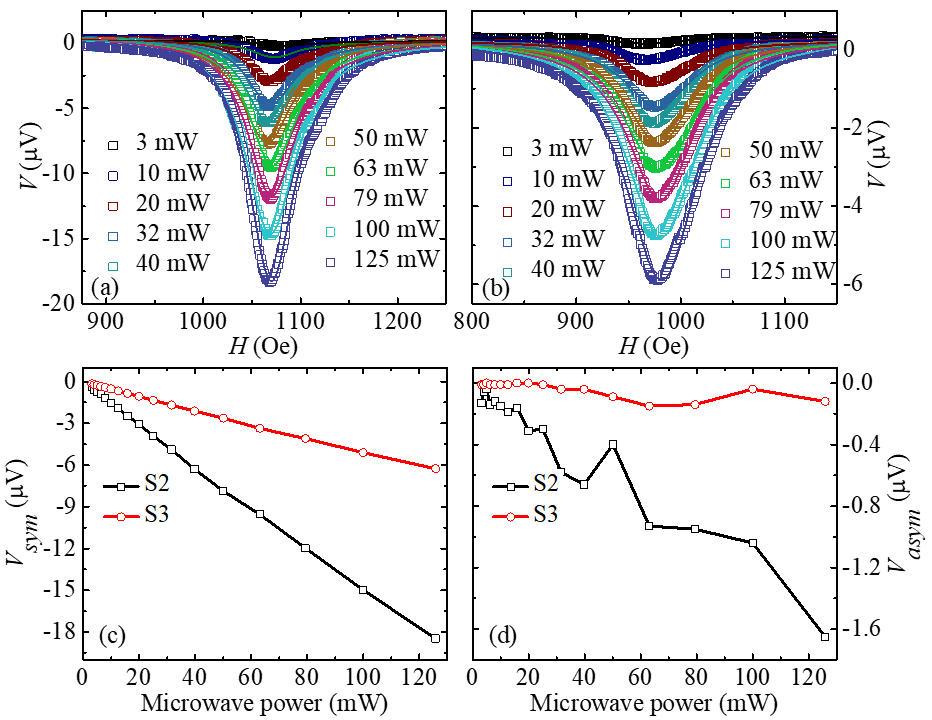}
	\caption{(a) and (b) Power dependent voltage signal for samples S2 and S3 measured at a $f$ = 7 GHz. (c) and (d) show power dependent $V_{sym}$ and $V_{asym}$ components for samples S2 and S3, respectively}
	\label{fig:figure-5}
\end{figure}

In order to further confirm that the $V_{meas}$ is primarily  due to spin pumping, we have performed microwave power dependent ISHE at 7 GHz. Power dependent measurement was  performed in microwave power range of 3 to 125 mW. Microwave power dependent voltage signal is shown in Fig.~\ref{fig:figure-5}(a) and 5(b) for samples S2 and S3, respectively. Figure~\ref{fig:figure-5}(c) show the power dependent symmetric part of voltage for samples S2 and S3. The linear increase in microwave power leads to increase in $V_{sym}$ signal strength for both the samples S2 and S3, which confirms that $V_{meas}$ was mainly due to spin pumping. Figure~\ref{fig:figure-5}(d) shows the $V_{asym}$ dependency over microwave power for samples S2 and S3.

\section{\label{sec:level4}CONCLUSIONS:\protect} 

\begin{table*}[ht]
\caption{{Comparison of various parameters from literature for LSMO/Pt bilayers.} }
\centering
\begin{tabular}{ccccccc}
\hline
Authors & System & Preparation technique & $V_{sp}$ ($\mu V$) & $g_{eff}^{\uparrow\downarrow}$ ($m^{-2}$) & Power (mW) & $\alpha$ ($10^{3}$)\\ [1ex] 
		\hline
		Atsarkin et al.\cite{atsarkin2016resonance} & LSMO(80 nm)/Pt(10 nm) & PLD & 0.56 & $10^{16}$ $\hyp{}$ $10^{17}$ & 250 & $\hyp{}$ \\    
		\hline
		Luo et al.\cite{luo2017spin} & LSMO(20 nm)/Pt(6$\hyp{}$30 nm) & PLD & 8 & 1.8$\times$ $10^{19}$ & 100 & 4-8 \\ 
		\hline
		Lee et al.\cite{lee2016magnetic} & LSMO(30 nm)/Pt(5$\hyp{}$9 nm) & PLD & 0.3 & 2.1$\times$ $10^{19}$ & $\hyp{}$ & 1.9$\hyp{}$2.9 \\ 
		\hline
		Luo et al.\cite{luo2012spin} & LSMO(29 nm)/Pt(10nm) & PLD & $\sim$1 & $\hyp{}$ & 100 & $\hyp{}$ \\ 
		\hline
		Luo et al.\cite{luo2015enhanced} & LSMO(20 nm)/Pt(5.5 nm) & PLD &$\sim$5 & $\hyp{}$ & 125 & $\sim$5.93 \\ 
		\hline
		Luo et al.\cite{luo2014influence} & LSMO(26)/Pt(5.5) & PLD & $\sim$3.25 & $\hyp{}$ & 40 & $\sim$6.50 \\ 
		\hline
		This work & LSMO(20nm)/Pt(3nm) & OMBE & 20.05 & 1.48$\times$ $10^{19}$ & 25 & 4.60 \\ 
		\hline
\end{tabular}
\end{table*}

We have studied spin pumping and ISHE for LSMO/Pt bilayer samples prepared by oxide molecular beam epitaxy. We have observed a decrease in the value of $\alpha$ with increase in the Pt thickness. This decrease in $\alpha$ value may be due to anti-damping like torque. At the low value of $\alpha$, we have observed high spin pumping voltage, which makes this system ideal for the development of power efficient spintronics devices. In Table II we show the comparison of various parameters from literature for LSMO/Pt bilayers. We found spin Hall angle value 0.033 for 3 nm Pt thickness which is in range of previously reported values. It seems that the oxide molecular beam epitaxy is a suitable technique to prepare high quality complex oxides. Further study of manganite based system can give the way to control the spin to charge conversion efficiency for the future applications.

\begin{acknowledgments}
We thank to Department of atomic energy for providing financial support. SB thanks DAAD for providing financial support as a guest scientist to carry out the sample preparation at FZ Juelich, Germany. BBS acknowledges DST for INSPIRE faculty fellowship. PG and KR acknowledge UGC  and CSIR for JRF fellowships, respectively.
\end{acknowledgments}

\bibliographystyle{apsrev4-2}
\bibliography{reference}

\end{document}